\documentclass{emulateapj}
\usepackage{graphicx}
\usepackage[english]{babel}
\usepackage{amssymb}
\usepackage{rotating}
\usepackage{natbib}
\usepackage[bookmarks,bookmarksnumbered,colorlinks=true, citecolor=blue, linkcolor=black]{hyperref}

\def\apjl {ApJL}
\def\apjs {ApJS}

\def\aap {A\&A}

\def\mpc{{\rm{Mpc}~h^{-1}}}
\def\Msol {\rm {M}_\odot}

\def\Kpc {\rm Kpc}
\def\kms {\rm{km~s^{-1}}}
\def\Gyr {\rm Gyr}

\slugcomment{To be submitted to ApJ}

\shorttitle{Intra-cluster Globular Clusters}
\shortauthors{Ramos et al.}

\begin{document}

\title{Intra-cluster Globular Clusters in a Simulated Galaxy Cluster}

\author{Felipe Ramos-Almendares\altaffilmark{1,2}, Mario G. Abadi\altaffilmark{1,2},
Hern\'an Muriel\altaffilmark{1,2} and Valeria Coenda\altaffilmark{1,2}}

\altaffiltext{1}{Universidad Nacional de C\'ordoba, Observatorio Astron\'omico de C\'ordoba, C\'ordoba, Argentina}
\altaffiltext{2}{CONICET-Universidad Nacional de C\'ordoba, Instituto de Astronom\'ia Te\'orica y Experimental, C\'ordoba, Argentina}

\email{f.ramos.almendares@gmail.com}

\begin{abstract}
Using a cosmological dark matter simulation of a galaxy-cluster halo, we follow the temporal evolution of its globular cluster population.
To mimic the red and blue globular cluster populations, we select at high redshift $(z\sim 1)$ two sets of particles from individual galactic halos constrained by the fact that, at redshift $z=0$, they have density profiles similar to observed ones.
At redshift $z=0$, approximately 60\% of our selected globular clusters were removed from their original halos building up the intra-cluster globular cluster population, while the remaining 40\% are still gravitationally bound to their original galactic halos.
Since the blue population is more extended than the red one, 
the intra-cluster globular cluster population is dominated by blue globular clusters, with a relative fraction that grows from 
60\% at redshift $z=0$ up to 83\% for redshift $z\sim 2$.
In agreement with observational results for the Virgo galaxy cluster, the blue intra-cluster globular cluster population is more spatially extended than the red one, pointing to a tidally disrupted origin.
\end{abstract}



\section{Introduction} \label{sec:intro}

Galaxy clusters are extreme environments in which galaxies are subject to both strong mutual interactions and intense tidal force generated by the  central gravitational potential of the cluster.
These processes have a dramatic impact on galaxy morphology and their intrinsic properties such as stellar mass, gas content, star formation rate, colors, etc. (\citealt{Dressler:1980}, \citealt{Goto:2003}, \citealt{Coenda:2006}, \citealt{Coenda:2009}). Tidal stripping removes mass from individual galaxies, which becomes part of the intra-cluster medium as it is bound to the  central potential of the cluster \citep{Moore:1999}.

Diverse observational evidence points to the presence of stellar material removed from galaxies in the intra-cluster medium. 
\cite{durrell_intracluster_2002} identified red giant star candidates in the intra-cluster medium of the Virgo galaxy cluster. In nearby galaxy clusters, there is evidence of the presence of planetary nebulae in the intra-cluster medium \citep{mendez_more_1997,gerhard_kinematics_2007, Castro-rodriguez:2009,Longobardi:2013}, novae \citep{Neill:2005} and supernovae \citep{GalYam:2003,Sand:2011,Graham:2015}.
Stars that have been removed from their parent galaxy can be observed in galaxy clusters as a diffuse light, called intra-cluster light. 
However, its observational detection is very difficult due to the extremely low surface brightness of the intra-cluster light, which contains $\sim 10\% - 30\%$ of the total optical luminosity of the galaxy clusters \citep{gregg_galaxy_1998,feldmeier_deep_2002,mihos_diffuse_2005, Krick:2007, Burke:2012, Mihos:2015}. 

Globular cluster systems are one of the more extended baryonic components of galaxies, so it is expected that these objects will be especially prone to being removed by tidal stripping. \cite{white_globular_1987}, was the first author to hypothesize the existence of intra-cluster globular clusters; i.e. globular clusters that are no longer bound to individual galaxies. \citet{West:1995} suggested that a large intra-cluster globular cluster population could explain the high specific frequency $S_N$ values for large ellipticals in the centers of clusters. A small number of intra-cluster globular clusters have been confirmed in the Virgo ($N=4$) \citep{williams_virgos_2007} and Fornax galaxy clusters ($N=\sim 75$) \citep{Bassino:2003, Bergond:2007, Schuberth:2008}. \citet{coenda_tidal_2009} analyzed the $S_N$  of early type galaxies in Virgo and
found that  larger values  are at greater cluster-centric
distances. They suggested that galaxies closer to the Virgo center are
losing a significant fraction of their globular clusters due to tidal stripping.
The first statistically significant detection of intra-cluster globular clusters was carried out by \cite{lee_detection_2010}, estimating that there are about $\sim 11900$ intra-cluster globular clusters in Virgo. \cite{peng_hst/acs_2011} identified globular clusters in the Coma galaxy cluster, finding $\sim 47000$ intra-cluster globular clusters inside a radius of $520 \Kpc$ $(\sim 0.2~r/r_{vir})$ from its center. Those authors estimated that about 30\% to 40\% of globular clusters in Coma are intra-cluster globular clusters.  \cite{west_globular_2011} confirmed the presence of $\sim 1300$ globular clusters in the Abell 1185 galaxy cluster that apparently are not bound to any galaxy.
Recently, \citet{Harris:2017} estimate $\sim 50000-80000$ intra-cluster globular clusters in the Perseus cluster, with a radial distribution extending several kiloparsecs away from the cluster center. 

The are several studies in slightly more distant clusters. \citet{alamo-martinez_rich_2013} identified $\sim 80000$ intra-cluster globular clusters within the central $400 \Kpc$ in Abell 1689.
\cite{dabrusco_extended_2016} found an asymmetric structure that links the globular cluster system of NGC 1399 with near galaxies in the Fornax galaxy cluster. \cite{lee_globular_2016} identified globular clusters and Ultra Compact Dwarf (UCD) galaxies in Abell 2744, finding $\sim 1.7\times 10^{6}$ intra-cluster globular clusters. Those authors found that intra-cluster globular clusters have a number density profile steeper than the dark matter profile.

\begin{figure*}
	\includegraphics[width=16cm]{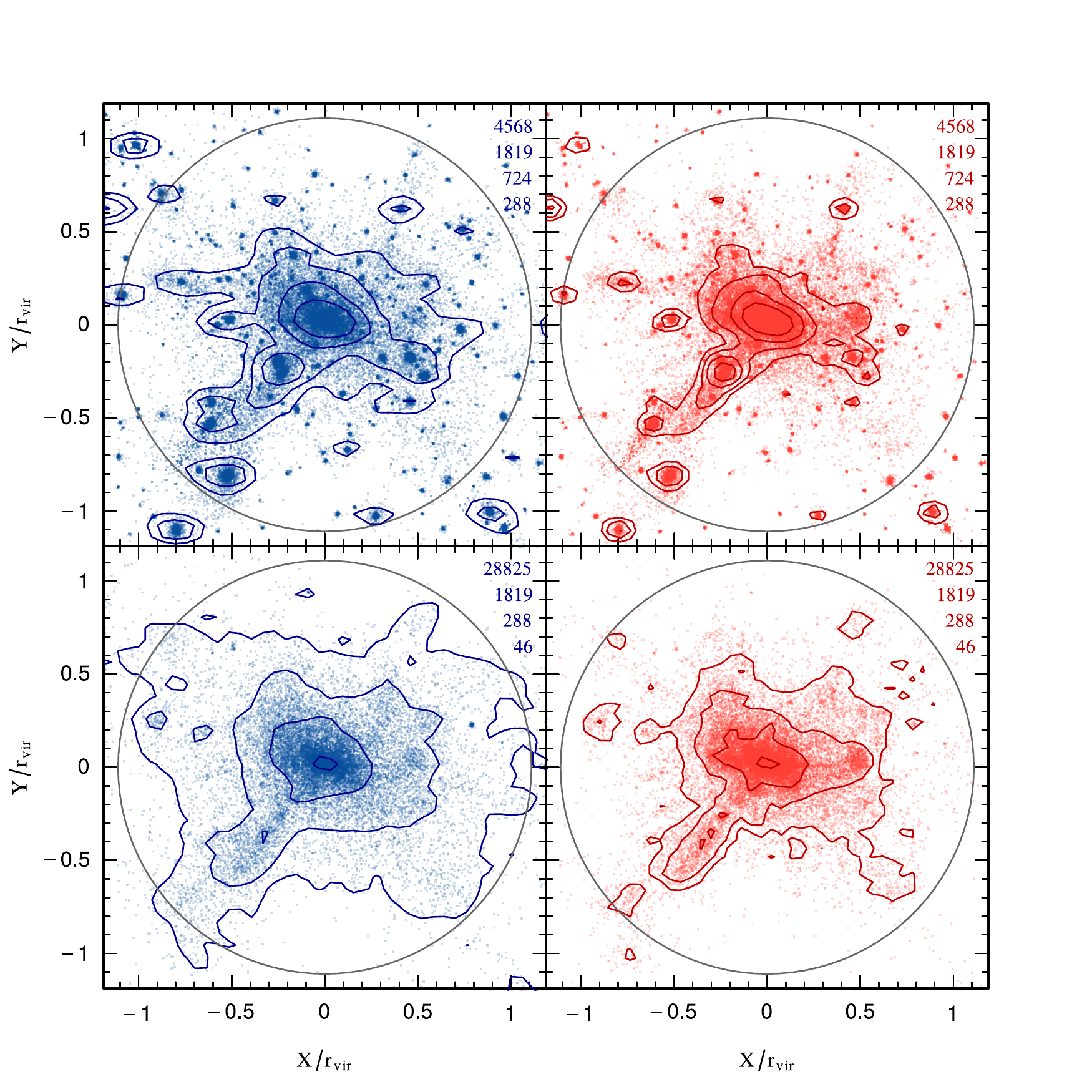}
	\caption{Projected spatial distribution of dark matter particles selected as blue (left panel) and red (right panel) globular clusters at redshift $z=0$. Upper panels correspond to all globular clusters while bottom panels correspond only to intra-cluster globular clusters. Colored curves are isocontours of projected number density with values given in the upper right corner of each panel. Solid black circle shows the galaxy cluster virial radius $r_{vir}=1.1 \mpc$}
	\label{fig:den}
\end{figure*}

From the theoretical point of view, \cite{YB:2005} used a cosmological N-body numerical simulation and randomly selected dark matter particles as globular clusters inside galactic halos. They found that nearly 30\% of globular clusters inside a galaxy cluster are intra-cluster globular clusters and that a small fraction of globular clusters removed from their halo can be found outside the virial radius of the galaxy cluster (inter-cluster globular clusters). \cite{bekki_spatial_2006} obtained similar results, finding that between 20\% and 40\% of total globular clusters inside a galaxy cluster are intra-cluster globular clusters with a density profile that is steeper than the dark matter profile. In \citet{ramos_tidal_2015}, we used a dark matter-only numerical simulation to study tidal stripping of globular clusters in galaxies orbiting a galaxy cluster. We found that halos lose on average 29\% (16\%) of blue (red) globular clusters, and that the fraction of globular clusters that are removed from an individual halo depends strongly on the orbital trajectory.

The intra-cluster globular clusters are mostly metal-poor, with colors typical of globular clusters of dwarf galaxies, and they are found even far from massive galaxies \citep{lee_detection_2010,peng_hst/acs_2011,alamo-martinez_rich_2013, Durrell:2014}. \citet{Bekki:2009} argue that intra-cluster globular clusters are formed as a result of tidal stripping of globular clusters, initially within galaxy-scale halos during hierarchical growth of clusters via halo merging.  
\citet{lee_detection_2010} suggest a mixture scenario for the origin of globular clusters: metal-poor globular clusters would form in low-mass dwarf galaxies and metal-rich globular clusters would form later in massive galaxies or in dissipational merging galaxies. Alternatively, several authors suggest that intra-cluster globular clusters were formed in intra-cluster medium through gravitational instability of primordial density fluctuations, without ever residing in galaxies \citet{YB:2005,Griffen:2010}. Another possibility about the origin of intra-cluster globular clusters is that they form in thermal instabilities in cooling flows and gas-rich environments \citep{Cen:2001,Griffen:2010}.

\begin{figure}
\includegraphics[scale=0.35]{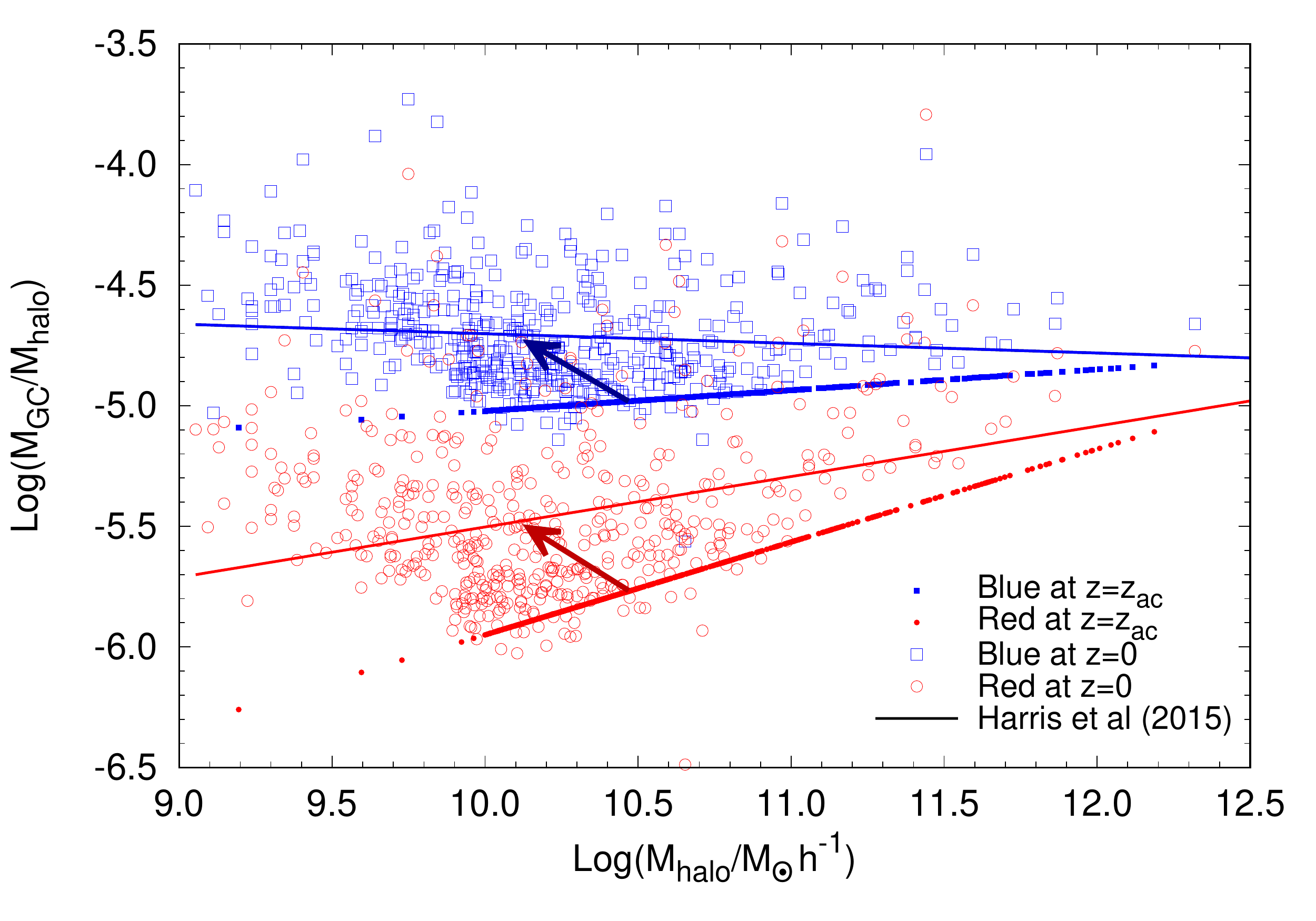}
\caption{Correlation between the ratio of globular cluster mass to its host subhalo mass as a function of the subhalo mass at selection time $t_{ac}$ (solid symbols) and at redshift $z=0$ (open symbols). Blue (squares) and red (circles) colors correspond to blue and red globular clusters, respectively. Solid lines 
show the power law fit obtained by \citet{harris_dark_2015} at $z=0$; our selection method is calibrated in order to reproduce approximately these observational trends. Arrows show the median shift of an individual globular cluster in our sample from accretion time $t_{ac}$ to redshift $z=0$.}
\label{fig:masa}
\end{figure}
Using a cosmological N-body numerical simulation in the framework of the $\Lambda$ Cold Dark Matter model, we study the blue and red intra-cluster globular cluster populations in a Virgo-like galaxy cluster. 
The aim of this work is to test whether the loss of globular clusters by gravitational interactions can explain the observed population of intra-cluster globular clusters.
In section \ref{sec:methods} we describe the methods that we applied in order to mimic the globular cluster population. In section \ref{sec:results} we present the main results about intra-cluster globular cluster populations. Finally in \ref{sec:conclusions} we summarize the results and conclusions.

\section{Methods}\label{sec:methods}
\subsection{Numerical Simulation}

As in \citet{ramos_tidal_2015}, we use a dark matter-only cosmological numerical simulation that follows the formation of a 
Virgo-like galaxy cluster-sized halo (the halo labeled "h14" in Table 1 of \citet{ludlow_secondary_2010}). Briefly, the 
simulation was performed using the GADGET2 code \citep{springel_cosmological_2005} in a cosmological box of 100 
$\mpc$ comoving on a side, assuming the following cosmological parameters: $H_0=73 \kms \mpc$, $\Omega_0=0.25$ and $\Omega_\Lambda=0.75$. Then, the region was resimulated at a higher resolution using the zoom-in technique of \citet{klypin_resolving_2001}, 
which has $1.44\times 10^7$ dark matter particles with masses of $5.4 \times 10^7 \Msol h^{-1}$ and a gravitational 
softening of $1.5 \Kpc~h^{-1}$. At redshift $z=0$ the halo has a virial mass $M_{vir}=1.71\times 10^{14} \Msol h^{-1}$, 
which corresponds to a virial radius of $r_{vir}=1.1 \mpc$; i.e. the radius where the inner mass density is 200 times 
the critical mass density of the universe. There are $3.16\times 10^6$ dark matter particles within the virial radius 
and its shape is fairly close to spherical with normalized semi-axis ratios given by $b/a=0.87$ and $c/a=0.83$.
Using the SUBFIND algorithm \citep{springel_populating_2001}, \cite{ludlow_secondary_2010} selected the main halos 
and their respective subhalos in all 100 snapshots, equally spaced in $\log(a)$ (here $a$ is the expansion factor of the universe), 
from redshift $z=19$ to $z=0$. Further technical details can be found in \cite{ludlow_secondary_2010}.

\begin{figure}
\centering
	\includegraphics[width=9cm]{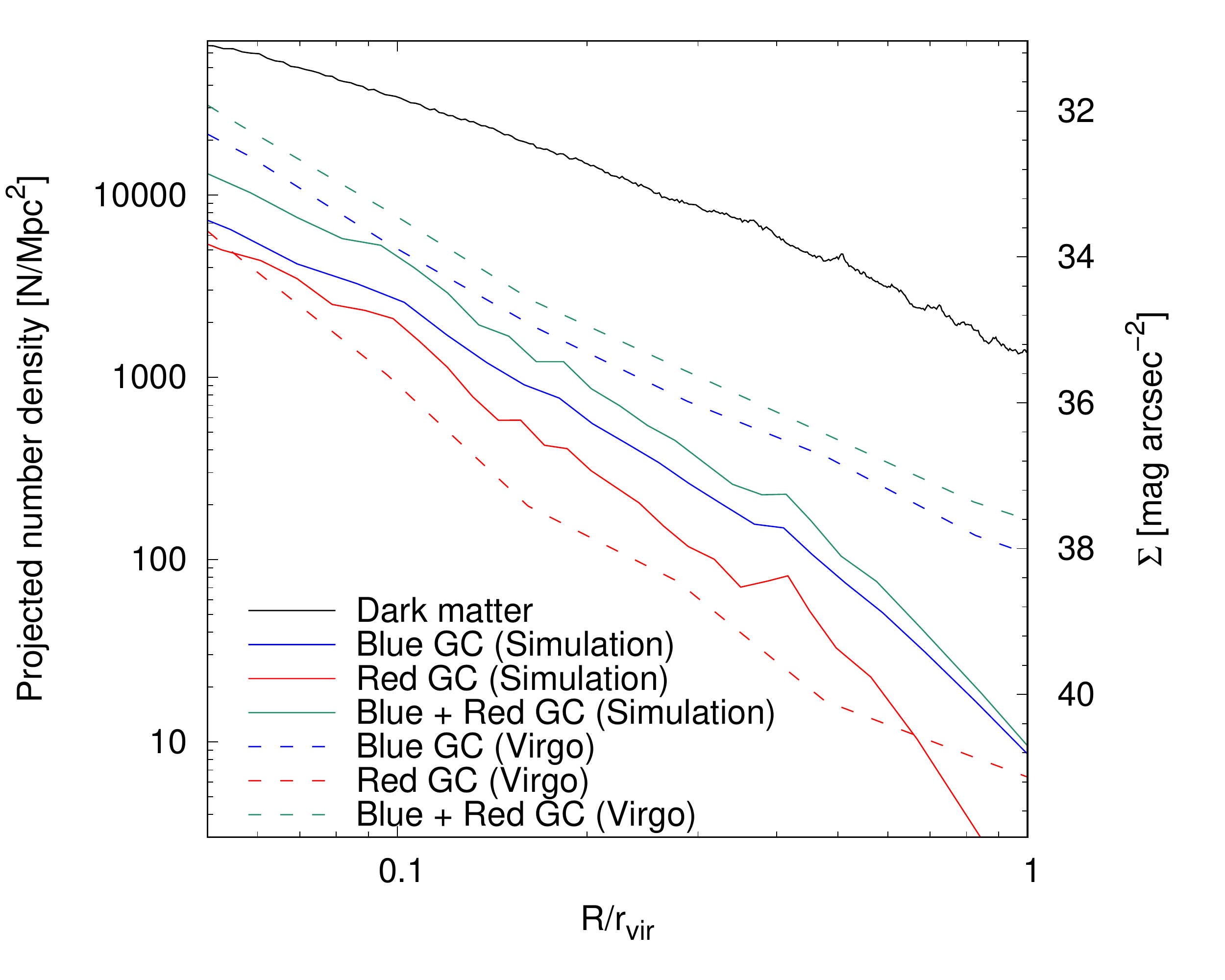}
	\caption{Projected number density of intra-cluster globular clusters as a function of the projected 
	cluster-centric distance $R$ normalized to the galaxy cluster virial radius $R_{vir}$. Solid blue and red 
	lines correspond to blue and red intra-cluster globular cluster population, respectively, in our simulation. 
	Dashed blue and red lines correspond to intra-cluster globular clusters in the Virgo cluster measured by 
	\citet{lee_detection_2010}. Analogously, green lines show blue plus red intra-cluster globular clusters in our simulation (solid) and
	in the Virgo cluster (dashed). For comparison, the solid black line shows the projected number density 
	of the main dark matter halo (removing substructures) in arbitrary units.}
	\label{fig:figperf}
\end{figure}

\subsection{Globular cluster selection}

We select all dark matter subhalos accreted by the central halo that are
resolved by at least 200 particles at accretion time $t_{ac}$, defined as the time when its
cluster-centric distance is typically $\approx 1.3 r_{vir}$.
We follow the temporal evolution of
each subhalo, computing its mass, density profile and orbital evolution in the galaxy cluster potential.
Our sample comprises 625 galactic subhalos  that at $t_{ac}$ have virial masses 
ranging between $1.08 \times 10^{10} \Msol h^{-1}$ and $6.01 \times 10^{12} \Msol h^{-1}$ with a median of $4.0 \times 10^{10} \Msol h^{-1}$.
Subhalos are continuously accreted from the beginning of the simulation to the present time with a median 
accretion time of 
$6.5\Gyr$ which corresponds to redshift $z \sim 0.82$.
As in \citet{ramos_tidal_2015}, we follow the method described by \cite{bullock_tracing_2005} and \cite{penarrubia_tidal_2008}. Briefly, we assume that the dark matter density
profile of each subhalo in our sample can be approximated by a Navarro-Frenk-White (NFW) fitting equation  \citep{navarro_structure_1996}:
\begin{equation}
\rho_{NFW}=\frac{\rho_{NFW}^{0}}{(r/r_{NFW})(1+r/r_{NFW})^2}
\end{equation}
where $\rho_{NFW}^0$ is the characteristic density and $r_{NFW}$ is the scale length.
To mimic the red and blue globular cluster system in each galactic subhalo, we select at accretion time
$t_{ac}$ dark matter particles as tracers of the corresponding globular cluster populations with
a \cite{hernquist_analytical_1990} density profile:

\begin{equation}
\rho_{H}=\frac{\rho_{H}^{0}}{(r/r_{H})(1+r/r_{H})^3}
\end{equation}
where $\rho_H$ and $r_H$ are a characteristic density and scale length, respectively.
The corresponding energy distribution function are given by:
\begin{equation}
f(\epsilon)=\frac{1}{8\pi}\Big[\int_0^{\epsilon} \frac{d^2\rho}{d\psi^2}\frac{d\psi}{\sqrt{\epsilon-\psi}}+
            \frac{1}{\sqrt{\epsilon}}\Big(\frac{d\rho}{d\psi}\Big)_{\psi=0}\Big]
\end{equation}
where $\epsilon$ is the relative specific energy and $\psi$ is the relative gravitational potential.
\begin{figure}
\centering
    \includegraphics[width=9cm]{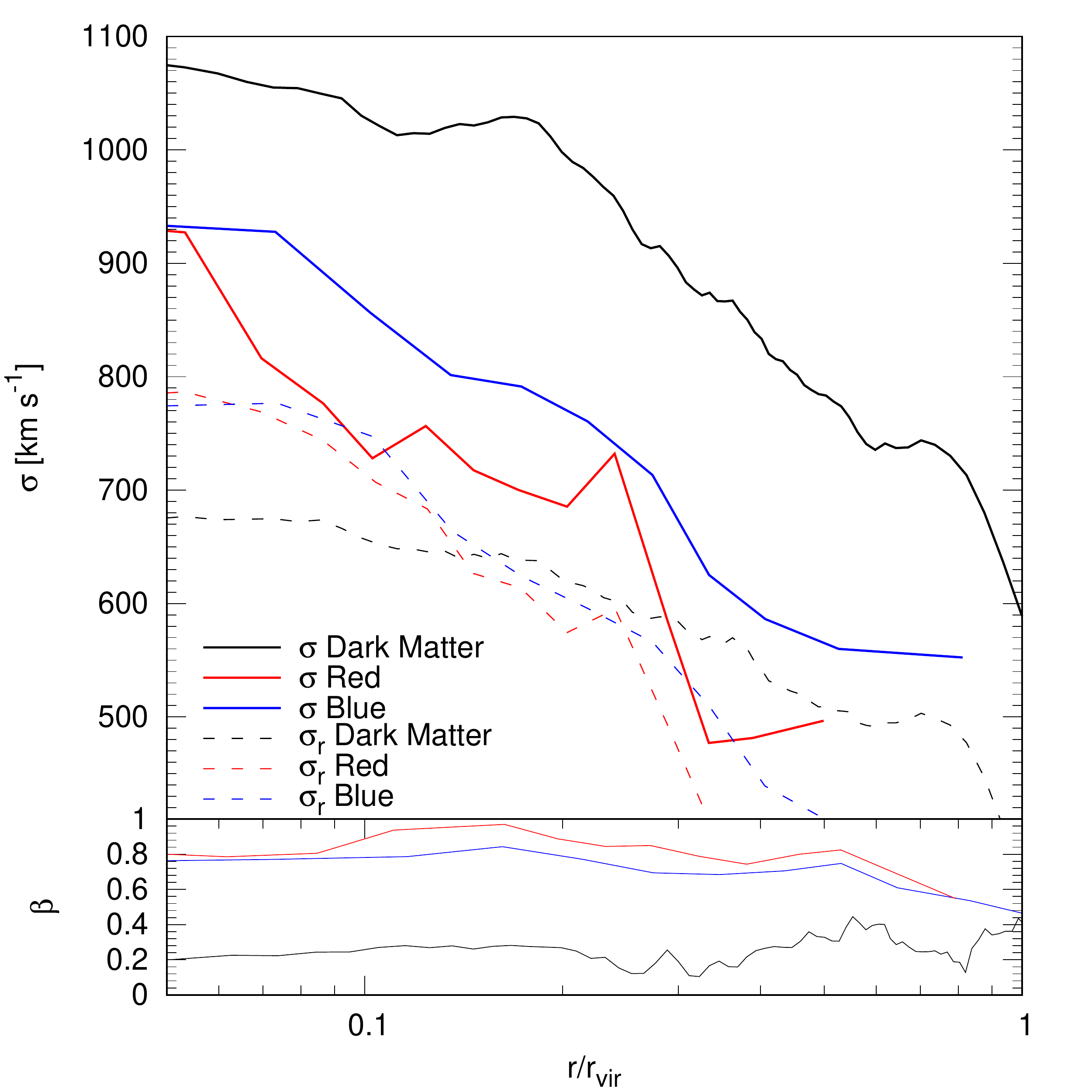}
    \caption{Solid (dashed) lines show total (radial) velocity dispersion as function of cluster-centric
    distance $r$ normalized to the virial radius $r_{vir}$ at redshift $z=0$. Blue and red color
    corresponds to blue and red intra-cluster globular clusters, respectively.
    Black lines correspond to the main dark matter halo removing substructures.}
    \label{fig:figdv}
\end{figure}
The method outlined by \cite{bullock_tracing_2005} selects dark matter particles in bins
$\epsilon + \Delta \epsilon$  according to a probability given by the ratio
$f_H(\epsilon)/f_{NFW}(\epsilon)$, where the subscripts $NFW$ and $H$ refer to a NFW and Hernquist density profile, respectively.
Following \cite{bekki_spatial_2006},
we truncate our selection of dark matter particles up to a radius $r_{cut}=r_{half}/3$,
where $r_{half}$ is the half mass radius of the dark matter subhalo. This truncation radius is imposed in order
to take into account that in general the baryonic components, including the globular cluster system, are
less extended than the dark matter subhalo.

 We have fitted a NFW profile for each subhalo at its accretion time $t_{ac}$ finding 
scale length parameters $r_{NFW}$ that range between $1.5\Kpc~h^{-1}$ and $92.6 \Kpc~h^{-1}$, with a median of
$8.7\Kpc~h^{-1}$. We have assumed a linear relation between the NFW and Hernquist scale length, $r_H = \gamma r_{NFW}$ with $\gamma$ values chosen to give at
redshift $z=0$ a projected number density profile similar to the power laws 
observed for red and blue populations. We adjusted $\gamma$ following the
results of \citet{coenda_tidal_2009} for the Virgo cluster which find 
logarithmic slopes of $-2.4$ and $-1.7$ for red and blue globular clusters, respectively. Applying this constrains 
we obtain $\gamma = 3$ for blue and $\gamma = 0.5$ for red globular cluster populations. 

Upper panels in Figure \ref{fig:den} show the projected spatial distribution, at redshift $z=0$, of all
dark matter particles selected at accretion time, $t_{ac}$, as blue ({left panel}) and red (right panel) globular clusters.
Lower panels correspond only to particles removed from their original galactic subhalos that at redshift $z=0$ are part of the intra-cluster medium.
As in \citet{ramos_tidal_2015}, we use a spatial criterion to determine if a
globular cluster particle is still associated to its subhalo or if it was removed. We consider that a particle
was removed from it subhalo when its distance to the subhalo center is greater than the distance of the farthest
particle identified by SUBFIND algorithm and by the following two snapshots and also at $z=0$. Particles selected
as globular clusters and removed from their original subhalos build up the intra-cluster globular clusters population.
From a total of 625 subhalos tagged at $t_{ac}$, there are 488 identified as self bound structures while 137
merged with the central galaxy cluster at redshift $z=0$.

\begin{figure}
\centering
    \includegraphics[width=9cm]{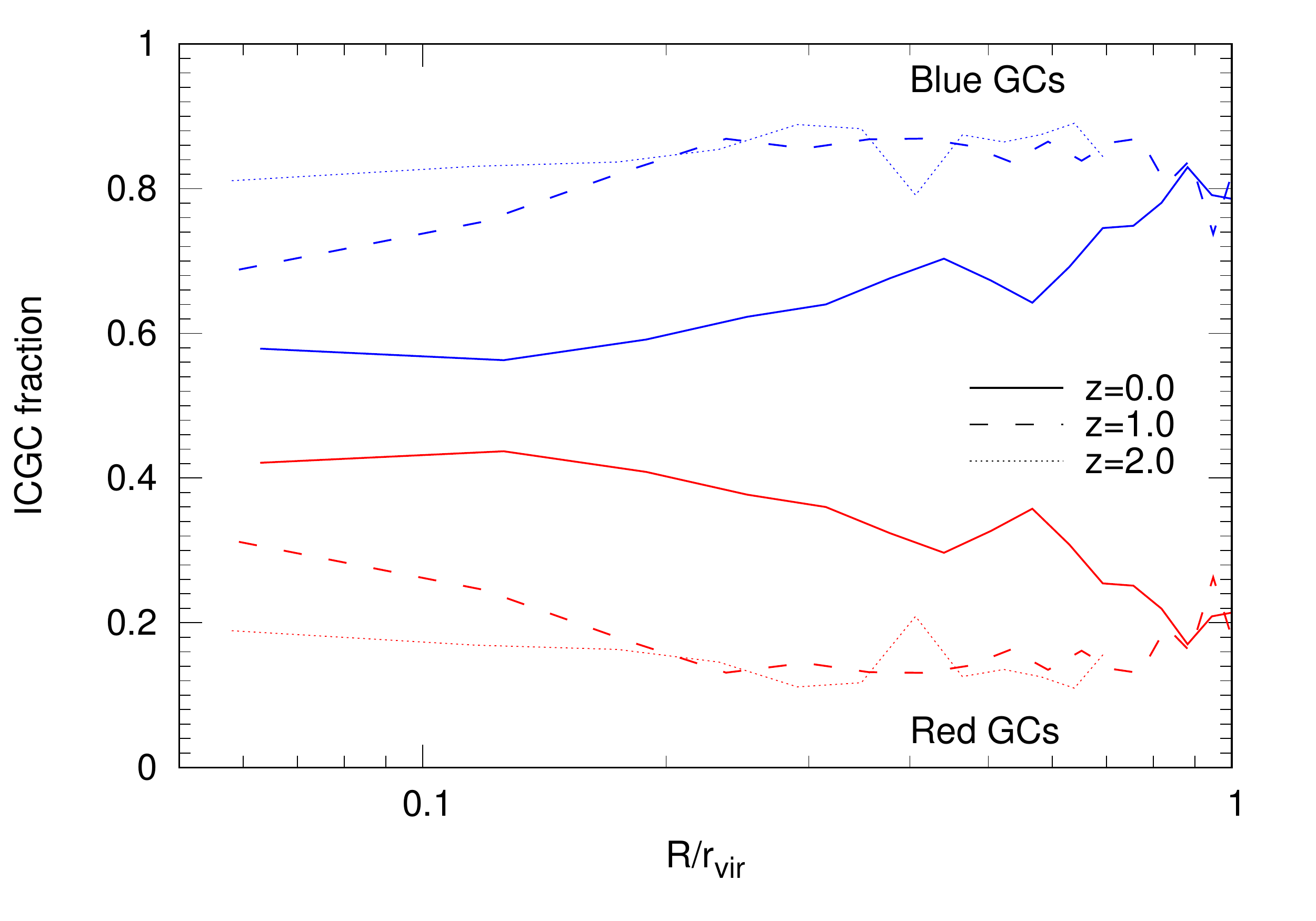}
    \caption{Fraction of blue (blue lines) and red (red lines) intra-cluster globular clusters as
    a function of the normalized cluster-centric distance $R/r_{vir}$ for different redshifts:
    solid lines correspond to $z=0$, dashed lines to $z=1$ and dotted lines to $z=2$.}
    \label{fig:figfaz}
\end{figure}

\subsection{Globular cluster mass}

In order to assign a globular cluster mass to each individual dark matter particle selected as a tracer 
of the globular clusters population, we use the observational results described by \cite{harris_dark_2015}. Using a 
sample of 419 galaxies (\citealt{harris_catalog_2013}), they find a correlation between the total halo
mass of a galaxy $M_{halo}$ and its total mass in globular cluster $M_{GC}$ for blue and red systems. 
They provide a power law fitting equation that can be written as: $\log(M_{GC}/M_{halo})= \alpha + 
\beta \log(M_{halo}/\Msol h^{-1}) $ with $\alpha=(-4.32,-7.62)$ and $\beta=(0.04,0.21)$ for (blue, red) globular clusters.
At selection time $t_{ac}$, we assume that the relation between $M_{GC}$ and $M_{halo}$ follows a power law 
(without any scatter), with parameters adjusted to reproduce, at redshift z=0, the observational trends 
obtained by \cite{harris_dark_2015}. 
For 488 subhalos that are identified as self-bound structures by 
SUBFIND \citep{springel_populating_2001}, we show in Figure \ref{fig:masa}, the ratio $M_{GC}/M_{halo}$ 
between globular cluster mass and dark matter halo mass as a function of $M_{halo}$.

\begin{figure}[htb!]
\centering
\[\includegraphics[width=9cm]{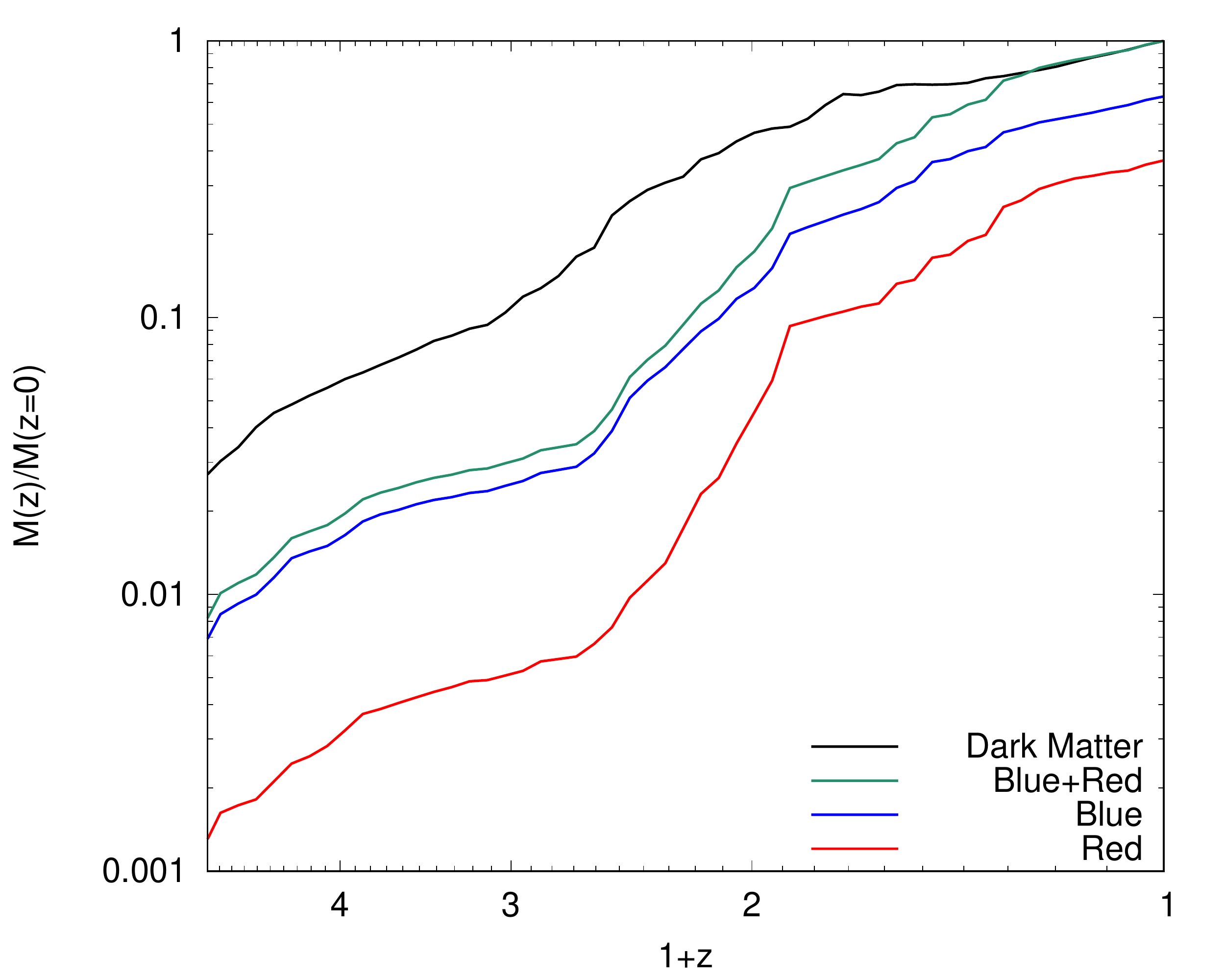}\]
\caption{Temporal evolution of intra-cluster globular cluster mass $M(z)$ normalized to the total 
	intra-cluster globular cluster mass at redshift $z=0$, $M(z=0)$ for blue (blue lines), red (red lines) 
	and blue plus red intra-cluster globular cluster populations (green lines). For comparison, the solid black line 
	shows the temporal evolution of the galaxy cluster dark matter halo normalized to its virial mass at redshift $z=0$.}
	\label{fig:figmvt}
\end{figure}

The correlation is shown both at redshift $z=0$ (open symbols) and at accretion time $t_{tac}$ (solid symbols). 
Solid lines show the correlation obtained by \cite{harris_dark_2015} which fits the relation for redshift $z=0$
(open squares). Arrows show the median loss of a subhalo mass $M_{halo}$ and the relative increase in the ratio 
$M_{GC}/M_{halo}$ from selection time $t_{ac}$ to redshift $z=0$. 
At redshift $z=0$, galactic subhalos have a mean loss of 55\% of their original mass at 
accretion time $t_{ac}$, while intra-cluster globular clusters have a loss of only 18\%. 
Thus the ratio $M_{GC}/M_{halo}$ increases by a factor of 1.8, as shown by the arrows 
in Figure \ref{fig:masa}. This probably reflects the fact that intra-cluster globular clusters have a more concentrated density profile than dark matter and thus less prone to tidal removal.
On average subhalos lose $\approx$ 16\% and 29\%
of their blue and red globular cluster population, respectively \citep{ramos_tidal_2015}.

It should be stressed that our model does not include the effect of globular cluster destruction: each dark matter particle selected as a globular cluster at high redshift remains intact to redshift $z=0$ keeping constant the total number of globular clusters. Our model only takes into account changes in the ratio 
between the number of globular clusters in galaxies and in the intra-cluster medium. 
Globular clusters can lose mass, and eventually be completely dissolved, by several effects such as stellar evolution (ejecta, stellar winds or supernovae) or dynamical processes (escapes by two-body scattering, gravitational shocks or dynamical friction).
\cite{Fall2001} use analytical models to compute these effects finding that 
the globular cluster mass function can be seriously affected for masses below $2 \times 10^5\Msol$. Then, for the more massive globular clusters or those that 
stay away from strong gravitational field generated by the baryonic components of galaxies, mass loss is not probably a concern.
In any case, it should be taken into account that most of our predictions 
are an upper limit.

\section{Results}\label{sec:results}
\subsection{Number density of intra-cluster globular clusters}

In order to compare our numerical simulation predictions with observational 
results, we compute number density isocontours (see Figure \ref{fig:den}) and profiles 
(see Figure \ref{fig:figperf}), assuming a constant mean mass of $2 \times 10^5 \Msol$ \citep{brodie_extragalactic_2006}, 
for both blue and red globular clusters.
Figure \ref{fig:den} shows four isocontour levels which correspond, from inside 
to outside, to a projected number density $\Sigma$ in units of Number/Mpc$^2$
indicated in the labels shown in the upper right corner of each panel.
Number density isocontours of the globular clusters are smooth, similar to observational results found
by \citet{lee_detection_2010} for globular clusters in the Virgo galaxy cluster, and by 
\citet{lee_globular_2016} for globular clusters plus ultra compact dwarfs in Abell 2744. 
However, simulated isocontours also show the presence of a large tidal stream that corresponds to a massive subhalo,
which has recently been disrupted by the galaxy cluster gravitational potential 
(see the bottom-left part of each panel in Figure \ref{fig:den}). 
This structure is the result of the tidal disruption of the second most massive
subhalo ($M=5.4 \times 10^{12} \Msol h^{-1}$ at $z_{ac}=1.11$), 
which enters the galaxy cluster virial radius. This tidal stream arises
after the third pericentric passage of the subhalo at redshift $z \sim 0.2$.
No observational evidence of such tidal streams of intra-cluster globular clusters has been reported so far. 
However, a density structure in the spatial distribution of intra-cluster globular clusters connecting galaxies has
been reported in Fornax by \citet{bassino_large-scale_2006} and confirmed by \citet{dabrusco_extended_2016}. 
These authors suggest that the galaxy NGC 1399 has stripped 
globular clusters from the less massive galaxy NGC 1387 forming this structure. 
They also report other elongated structures, mostly formed by blue globular 
Clusters, whose origins could be due to stripping events.

In Figure \ref{fig:figperf}, we show the projected number density profile of intra-cluster 
globular clusters as a function of the projected cluster-centric distance $R$, normalized to 
the galaxy cluster virial radius $R_{vir}$. Blue, red and green lines show the profile of blue, 
red and blue plus red globular clusters, respectively, with solid lines corresponding to our simulation and
dashed lines corresponding to the observed globular clusters in the Virgo galaxy cluster 
\citep{lee_detection_2010}. The solid black line shows the dark matter density profile arbitrarily normalized.

As a test of our mass assignment scheme we run a Montecarlo procedure over $\alpha$ and $\beta$ parameters taking into account uncertainties given by \citet{harris_dark_2015} obtaining very stable results. Errors in the projected 
number density profiles presented in Figure \ref{fig:figperf} are less than 0.2 dex.

Blue intra-cluster globular clusters have a more extended spatial distribution than the red ones, 
analogous to the observational results of \cite{lee_detection_2010}. This may be explained by 
the extended spatial distribution of blue globular clusters in galaxies being more prone to tidal stripping from their 
subhalos and therefore, at a greater cluster-centric distance than red ones. The latter may be 
removed when the halos pass through the central region of the cluster, where tidal forces are larger. 

Figure \ref{fig:figperf} also shows that the projected number densities of both blue and red 
intra-cluster globular clusters have a much steeper profile than that of the dark matter. 
A similar result was found by \citet{bekki_spatial_2006} using numerical simulations and 
by \citet{lee_globular_2016} in Abell 2744. This result is consistent with the scenario 
in which galaxies that are approaching the central region of clusters suffer a significant 
loss of globular clusters due to tidal effects.

\subsection{Kinematics of intra-cluster globular clusters}

The main panel of Figure \ref{fig:figdv} shows the total velocity dispersion $\sigma$ as 
a function of cluster-centric distance $r$, normalized to the virial radius $r_{vir}$ for 
dark matter (solid black line), and blue and red intra-cluster globular clusters 
(solid blue and solid red lines, respectively). Dashed lines correspond to the 
radial component of the velocity dispersion $\sigma_r$ using the same color coding.
In the inner regions, dark matter particles have a total velocity dispersion of 
$\sim 1100~ \kms$ while that of intra-cluster globular clusters is $\sim 900 ~\kms$. 
In the outer regions, $\sigma$ drops up to $\sim 600 \kms$ for dark matter,
and to slightly lower values for intra-cluster globular clusters, $\sim 500 \kms$, 
with blue ones always having higher velocity dispersion than red ones.
This result agrees with \citet{Schuberth:2008}, who found higher velocity 
dispersion for blue intra-cluster globular clusters in the Fornax cluster.
The bottom panel shows the anisotropy parameter:
\begin{equation}
\beta(r)=1-\frac{\sigma_t^2(r)}{2\sigma_r^2(r)}
\end{equation}
with $\sigma_t$ and $\sigma_r$, the tangential and radial components of the velocity dispersion,
using the same color coding as the main panel. While dark matter has $\beta \sim 0.2$, indicating 
orbits that are only slightly radially biased, intra-cluster globular clusters have $\beta \sim 0.8$, 
pointing up orbits that are strongly supported by radial velocity dispersion.
These results suggest that intra-cluster globular clusters are removed from
their galactic subhalos when they cross the central regions of the galaxy cluster. Since they are preferentially on radial orbits, they are prone to suffer strong tidal destruction. Given that our model does not include tidal destruction  our high value of the $\beta \sim 0.8$ anisotropy parameter should be considered only as an upper limit.

\begin{table}[h]
\begin{center}
\caption{Fraction of globular clusters}
\begin{footnotesize}
\begin{tabular}{crrr}
\hline \hline
 Subsample & N & \% Blue & \% Red \\
	(1) & (2) & (3) & (4) \\
  \hline
  Merged   & 137 &  65 & 79 \\
  Low mass  & 244 &  27 & 18 \\ 
  High mass & 244 &   8 &  3 \\
  \hline
\end{tabular}
\end{footnotesize}
\end{center}
\label{tab:por}
\footnotesize Note: Fraction of blue and red intra-cluster globular clusters according
to their host galactic subhalo mass at redshift $z=0$ in three different bins (column 1): 
those merged, low mass and high mass (see text for definition). 
Column 2 is the number of subhalos in each sample while columns 3 and 4 
are the relative fraction of blue and red, respectively.
\end{table}

\subsection{Evolution of intra-cluster globular clusters}

Figure \ref{fig:figfaz} shows blue and red intra-cluster globular clusters fractions as a
function of cluster-centric distance for three different redshifts 
($z=0$, $z=1$ and $z=2$) as indicated by the labels. 
In the inner regions, blue intra-cluster globular clusters represent nearly 60\% of the 
total intra-cluster globular cluster population and reach up to 80\%
in the outer parts.  
At higher redshifts, the fraction of blue intra-cluster globular clusters 
at large radii remains almost unchanged ($\sim$ 80\%) but increases from 
60\% to 80\% in
the inner regions. At redshift $z=2$, it is almost constant for all radii.

Figure \ref{fig:figmvt} shows the temporal evolution of the intra-cluster globular cluster mass, normalized to the total 
intra-cluster globular cluster mass of redshift $z=0$, for blue (blue lines), red (red lines), blue plus red (green lines) 
and dark matter (black lines). At redshift $z\sim 0.5$, the total mass in intra-cluster globular clusters
was $\sim$ 50\% of the final mass. Blue globular clusters dominate the intra-cluster globular cluster
population in the whole range of redshifts; approximately 60\% of the total intra-cluster globular 
cluster mass at redshift $z=0$ corresponds to the blue population.
The relative importance of blue intra-cluster globular clusters decreases with 
time from 80\% at redshift z$\sim$3 to 60\% at redshift $z=0$.

\subsection{Origin of intra-cluster globular cluster}

It has been proposed that blue intra-cluster globular clusters could be the result of globular 
clusters being stripped from low-mass dwarf galaxies, and red ones from 
giant elliptical galaxies \citep{lee_detection_2010}.
To test possible systematic effects of the origin of intra-cluster globular clusters, 
we analyzed their contribution by number, taking into account the subhalo mass at redshift $z=0$.
We defined three different bins in mass: high mass ($M_{halo}>1.38 \times 10^{10} \Msol h^{-1}$), 
low mass ($1.08 \times 10^{10} \Msol h^{-1}<$$~M_{halo}< 1.38 \times 10^{10} \Msol h^{-1}$), 
and merged ($M_{halo}>1.38 \times \Msol h^{-1}$). 
The mass limits between high and low bins correspond to the mean of our sample of 488 subhalos 
that survived at $z=0$ as self-bound structures, while the limit between low mass and merged corresponds 
to the limit detection of SUBFIND. Table \ref{tab:por} shows blue and red fractions of intra-cluster 
globular clusters from the three different bins in our sub-samples. 
By far, most of the intra-cluster globular clusters come from merged subhalos, with a higher fraction for red than blue ones. The second most important contribution corresponds to globular clusters stripped from massive halos (27 and 18\% for blue and red intra-cluster globular clusters, respectively) while low mass subhalos contribute only 8\% and 3\% for blue and red, respectively. We also analyzed the mass function of subhalos which, at $z=0$, are merged with the main galaxy cluster halo. We found that the fraction of merged subhalos is nearly independent of the subhalo mass at selection time $t_{ac}$. Since the number of globular clusters is proportional to the host subhalo mass, the main contribution to the intra-cluster globular clusters comes from high mass halos.

\section{Conclusions}\label{sec:conclusions}

We investigated the properties of blue and red populations of intra-cluster globular clusters, using 
a cosmological N-body numerical simulation that follows the formation and evolution of a dark matter 
halo which, at redshift $z=0$, has a mass similar to the Virgo galaxy cluster 
($M_{vir}=1.71\times 10^{14} \Msol h^{-1}$, $r_{vir}=1.1 \mpc$). We used the method outlined by
\cite{bullock_tracing_2005} and \cite{penarrubia_tidal_2008} to select blue and red globular 
cluster populations. The final sample comprises $625$ dark matter galactic subhalos that crossed the 
virial radius of the cluster with a virial mass greater than $M_{vir}\geq 1.08\times 10^{10} \Msol h^{-1}$.
We used observational results described by \cite{harris_dark_2015} to assign mass to the particles
selected as globular clusters. The results can be summarized as follows:

\begin{itemize}
\item At redshift $z=0$, we obtain a total intra-cluster globular cluster mass of 
$\sim 7\times 10^{8}\Msol h^{-1}$. In agreement with previous works, intra-cluster globular 
clusters are dominated by blue populations because they are more prone to be tidally stripped. 
At redshift $z=0$, approximately 60\% of the intra-cluster globular clusters are blue.

\item The projected numerical density of intra-cluster globular clusters clearly shows 
that blue intra-cluster globular clusters are more extended than red. Our results agree 
with observational results by \cite{lee_detection_2010} and \cite{lee_globular_2016} 
for Virgo and Abell 2744 clusters, respectively. 


\item We also found that the projected number density of both red and blue intra-cluster globular clusters have a much steeper profile than that of the dark matter component. This result agrees with the findings of \cite{lee_globular_2016} analyzing the globular cluster population in Abell 2744. Taking into account that globular clusters are a subset of dark matter particles selected to be a relatively more tightly bound than the dark matter as a whole, galaxies will need to get closer to the central region of clusters in order to lose their globular clusters due to tidal stripping.

\item At the cluster center, blue intra-cluster globular clusters represent almost 60\% of 
the total population of intra-cluster globular clusters. The different density profiles
of blue and red intra-cluster globular clusters mean that the fraction of 
blue intra-cluster globular clusters grows with the cluster-centric distanse, 
reaching 80\% in the outer parts. We found that this behavior strongly 
depends on redshift. At $z\gtrsim 1$, the same fraction is obtained at
$r=0.2r_{vir}$, indicating that the stripping of globular clusters starts with the blue population. At redshift $z \sim 0.5$, the total intra-cluster globular cluster mass is approximately half of that at $z=0$.  

\item At any cluster-centric distance, dark matter total velocity dispersion is systematically higher than that for intra-cluster globular clusters. While dark matter particles are preferentially supported by slightly radial orbits, intra-cluster globular clusters are basically orbiting in radially biased orbits.

\item We found that most of the intra-cluster globular clusters (65\% and 79\% for blue and red, respectively) 
come from halos that, at $z=0$, merged with the main galaxy cluster halo. The smallest contribution 
corresponds to low mass halos ($M_{halo}< 1.38 \times 10^{10}\Msol h^{-1} $ at $z=0$).  
We also found that the fraction of merged subhalos is nearly independent of the halo mass; 
consequently, massive halos are the main progenitors of the intra-cluster globular clusters. 
\end{itemize}

Our results are in good agreement with available observations of intra-cluster globular clusters in 
nearby galaxy clusters. According to our results, a Virgo-like galaxy cluster could be expected to
contain thousands of intra-cluster globular clusters, even at large distances from the cluster center. 
These results make intra-cluster globular clusters a powerful alternative to the study of 
diffuse intra-cluster light. We show that gravitational interactions (tidal stripping and/or 
destruction of halos) can completely explain the observed population of intra-cluster globular clusters.


\acknowledgments
We thank an anonymous referee for very useful comments that improved the original version of this paper. We thank Aaron Ludlow and Julio Navarro for making these
simulations available. This paper was partially supported by CONICET grants PIP 11220130100365CO and ANCyPT grants PICT 2012-1137, 
and grants from SECYT, Universidad Nacional de C\'ordoba, Argentina.

\bibliography{references.bib}


\end{document}